# Failure of the empirical OCT law in the $Bi_2Sr_2CuO_{6+\delta}$ compound

Z. Konstantinović[1], G. Le Bras[1], A. Forget[1], D. Colson[1], F. Jean[2], G. Collin[2], Z. Z. Li[3], H. Raffy[3], C. Ayache[1]

[1] *SPEC, DSM/DRECAM, CEA Saclay, 91191 Gif sur Yvette, France*
[2] *LLB, DSM/DRECAM, CEA-CNRS, CEA Saclay, 91191 Gif sur Yvette, France*
[3] *LPS, Bât 510, Université Paris-Sud, 91405 Orsay, Cédex, France*

**Abstract.** – We have studied the evolution of the thermoelectric power S(T) with oxygen doping of single-layered $Bi_2Sr_2CuO_{6+\delta}$ thin films and ceramics in the overall superconducting ($T_c$, $S_{290K}$) phase diagram. While the universal relation between the room-temperature thermopower $S_{290K}$ and the critical temperature is found to hold in the strongly overdoped region ($\delta > 0.14$), a strong violation is observed in the underdoped part of the phase diagram. The observed behaviour is compared with other cuprates and the different scenarios are discussed.

*Introduction.* – The determination of the number of holes injected in $CuO_2$ plane, *p*, which plays a key role in high-$T_c$ superconductivity, is still an open question. Universal empirical behaviour between the thermoelectric power (TEP) at 290 K, $S_{290K}$, and the ratio $T_c/T_{cmax}$, known as Obertelli-Cooper-Tallon law [1] (OCT), is currently used as a measure of *p*. This robust behaviour seems to be valid even in the case of $YBa_2Cu_3O_{7-\delta}$ where the charge is distributed between CuO chains and $CuO_2$ planes. The only known compound where this universal OCT law fails is the untypical $La_{2-x}Sr_xCuO_4$ (LSCO) superconductor [2]. Quite recently, however, a renewed interest has been raised by the observation of a departure from the OCT trend in the case of $Bi_2Sr_{2-z}La_zCuO_{6+\delta}$ (Bi(La)-2201) [3, 4].

Hall effect measurements can be also used to determine the hole number *p*, but in the case of the cuprates, the obtained values do not correspond to those expected from chemical determination [5]. Considering the structure, the average valence of Cu, denoted 2+*p*, leads in the stoichiometric $La_{2-x}Sr_xCuO_4$ compound to identify *p* with the concentration of Sr, x. In other cuprates, such as $Bi_2Sr_2Ca_{n-1}Cu_nO_{2n+4+\delta}$, with determined oxygen excess $\delta$, an estimation of p can also be obtained [6], although the charge transfer is more complicated due to the nonstoichiometricity of the structure. An alternative way to estimate *p* when the ratio $T_c/T_{cmax}$ is known is through a phenomenological law proposed by Presland *et al.* [7]. As pointed-out in Ref [3] and included in OCT behaviour itself [1], one can determine that the universal OCT relation holds if the hole number *p* estimated from phenomenological law, $p(T_c/T_{cmax})$, and from $S_{290K}$ [8], $p(S_{290K})$, lead to the same value.

Here, we present a detailed study on the doping evolution of the thermoelectric power in both $Bi_2Sr_2CuO_{6+\delta}$ thin films and ceramics in the overall superconducting region. The measurements on both type of samples appear to be essential to get complementary information. While in the case of ceramic samples we can determine the oxygen excess $\delta$, they remain intrinsically overdoped [9]. This restriction is overcome by using thin films which allow access to the strongly underdoped region of the ($T_c$, $\delta$) phase diagram, but without direct information about oxygen content in the sample. The observed properties on both types of materials are comparable in the overlap part of the phase diagram. An observed departure from the universal OCT behaviour is analysed and discussed.

*Experimental methods.* – The single-layered $Bi_2Sr_2CuO_{6+\delta}$ (Bi-2201) samples are either c-axis-oriented epitaxial thin films or polycrystalline samples.

Thin films were grown by RF magnetron sputtering on $SrTiO_3$ substrates [10]. The oxygen content of a given sample was changed by repeated annealing treatments in controlled atmosphere from an overdoped state with $T_c(R=0) \sim 4.5$ K (A1) to a strongly underdoped nonsuperconducting state with $T_c(R=0) \sim 0$ K (A6). The maximal critical temperature achieved was $T_{cmax}(R=0) \sim 16.5$ K (A2). The most highly overdoped sample with $T_c(R=0) \sim 0$ K (B) is obtained by an annealing treatment in an oxidising plasma. The doping evolution of the resistivity is given elsewhere [11]. The above described annealing procedure induces successive changes in the carrier content in the film in a controlled and reversible way [12].

Sintered samples were prepared using a classical solid reaction method [13]. Hole concentration can be adjusted through oxygen excess $\delta$, which has been quantitatively controlled by thermogravimetric techniques [9,14]. All polycrystalline samples have the same cationic composition. According to electron microprobe analysis, the resulting average cation ratio was found to be very close to Bi:Sr:Cu=2:2:1.

Thermoelectric measurements on thin films were performed using a conventional steady flow technique. The single-crystal substrate $SrTiO_3$ does not contribute to the measured thermopower. Temperature and voltage gradients were simultaneously measured using T-type (25 μm) thermocouples fixed on two gold sputtered contact pads. Detailed thermopower measurements on ceramic samples are published separately [15]. The observed TEP properties are fully reproducible in both thin films and ceramic samples. The critical temperature $T_c$ is determined from the onset of dc magnetisation [16, 17], which is found to be in good agreement with $T_c(R=0)$.

*Results and discussion.* – Typical temperature dependence of the thermoelectric power S(T) of $Bi_2Sr_2CuO_{6+\delta}$ thin films in the overall ($T_c$, $\delta$) phase diagram is shown in fig.1. The most overdoped (B) and the most underdoped (A6) states are situated at the limit of the superconducting region with $T_c \sim 0$, while A2 is in optimally doped state. The results are in good agreement with those obtained for Bi-2201 ceramics [15] with well-controlled oxygen excess $\delta$ [9]. The S(T) behaviour is strongly dependent on the oxygen doping. While a non-monotonous behaviour is observed for the underdoped states (A2-A6), a marked upward curvature is present in the overdoped states (B, A1). A possible scenario where S arises from two different drag and diffusion contributions was discussed previously [4,15,18].

In fig. 2 we show the variation of the critical temperature $T_c$ as a function of the room temperature thermopower $S_{290K}$. We observe good quantitative agreement between results obtained on thin films (closed symbols) and ceramic samples (open symbols). One may notice that in the overall superconducting region the room temperature thermopower remains mostly negative ($S_{290K}<0$). The optimally doped sample (A2) has $S_{290K}$ around -8 μV/K instead of the zero value expected from the OCT relation. A similar deviation from the universal relation was already reported in La-doped Bi-2201 single crystals [3], where $S_{290K}$ ~-6 μV/K was found in the case of optimally doped $Bi_2Sr_{1.7}La_{0.3}CuO_{6+\delta}$.

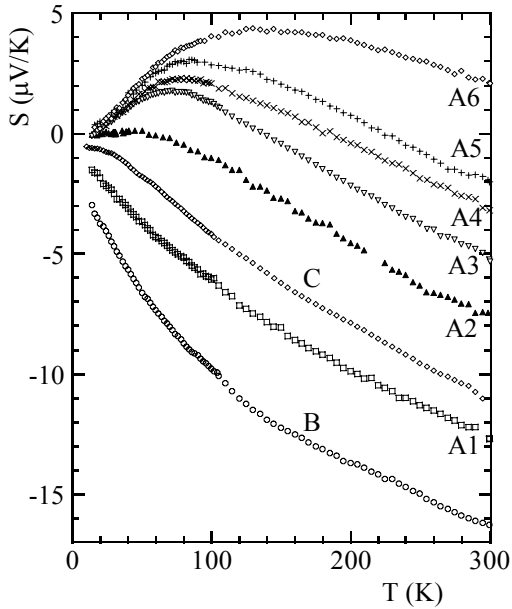 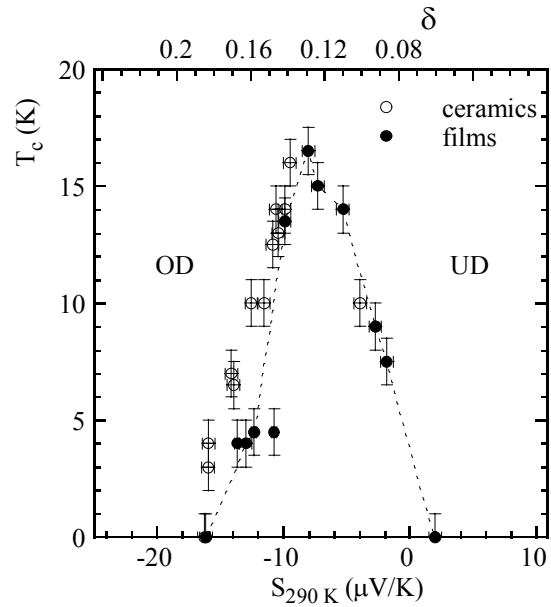

Fig. 1                                                                 Fig. 2

Fig. 1. – Temperature dependence of the thermoelectric power S for Bi-2201 thin films at different doping levels. The doping states, labelled by A1-A6, are obtained by successive annealing treatments on the same film. B and A6 are situated near the limit of superconducting region ($T_c$~0), while A2 is in an optimally doped state ($T_c$~16.5 K). C is an intermediary overdoped state ($T_c$~4.5 K).

Fig. 2. – Phase diagram ($T_c$, $S_{290K}$) of $Bi_2Sr_2CuO_{6+\delta}$ thin films (closed circles) and ceramics (open circles). The oxygen excess values δ, indicated in the top axis, are determined in the case of ceramic samples [9]. The overdoped (OD) and the underdoped (UD) region are also identify in the figure. Dotted line is just a guide to the eyes.

In fig. 3, $T_c/T_{cmax}$ is plotted as a function of $S_{290K}$ in order to show the disagreement between the universal behaviour (solid line) and our results obtained on non-substituted Bi-2201 (circles). The results reported on La-doped Bi-2201 [3] (dashed line) are also shown in the same figure. In comparison with OCT behaviour, the superconducting region ($T_c$,$S_{290K}$) in the Bi-2201 compound is extremely narrow. In the overdoped region, a good agreement is observed between these two single-layer Bi-2201 samples. Moreover, in the strongly overdoped region, the doping dependence of $T_c/T_{cmax}$ is very close to the universal behaviour. At the same time, in the underdoped part, a disagreement is observed between the La-doped Bi-2201 and non-substituted Bi-2201 single-layer, and in both cases the observed behaviour is very different from the universal one. A similar behaviour is also evidenced in the case of the singular branch in Bi(La)-2201 single crystals [4]. On the other hand, LSCO compounds shows completely different behaviour than that of Bi-2201. Over the entire superconducting range $S_{290K}$ remains positive, converging to the OCT behaviour in the strongly underdoped, non-superconducting part [2].

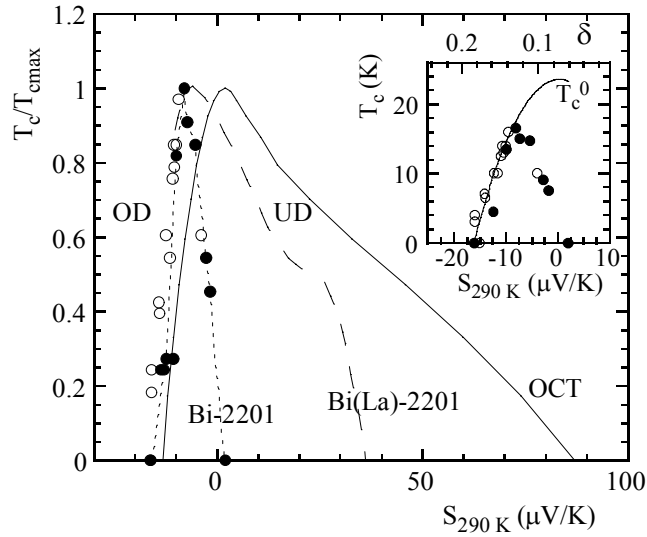

Fig. 3. – Comparison of $T_c/T_{cmax}$ vs. $S_{290K}$ between non-substituted Bi-2201 films (closed circles) and ceramics (open circles) and La-doped Bi-2201 single crystals (dashed line) [3]. Previously established empirical OCT relation is shown by solid line [1]. Inset: estimation of the fictitious $T_{cmax}^0$ value from the phenomenological law [7] with $p(S_{290K})$ [8] (solid line).

To quantify the agreement between different results, we compare the hole number $p$ determined from the phenomenological law $T_c/T_{cmax}=1-82.6(p-0.16)^2$ [7], $p(T_c/T_{cmax})$, with that from the OCT relation $S_{290K}=24.2-139p$ in the OD region [8], $p(S_{290K})$ (fig. 4). The agreement between $p(T_c/T_{cmax})$ and $p(S_{290K})$ is observed within 10% (dotted line) in the strongly overdoped region ($\delta>0.14$), while a disagreement between them is observed for $\delta<0.14$. As the two determined hole doping numbers are similar in the strongly overdoped region ($0.14<\delta<0.18$), the OCT relation holds within 10%, while it is strongly violated in the underdoped region. As it was discussed previously [3], the failure of the OCT behaviour could come from a somehow reduced either critical temperature or thermoelectric power. In the following we will discuss these two possibilities.

If we assume that measured $T_c$ values are somehow reduced from the real ones, without affecting the thermoelectric power, we can estimate the fictitious $T_c^0$ values from the phenomenological law [7] with the hole number determined from thermopower values $p(S_{290K})$ (inset of the fig. 3) [19]. In this case all measured samples (except A6) are situated in the overdoped region of the phase diagram. However this point remains questionable considering the general doping tendencies of the normal state properties of cuprates. The pseudogap opening in the underdoped region [20]. In our case, doping state A2 shows the linear T-variation of the resistivity, characterising the optimal doping state and the less-doped ceramic sample ($\delta\sim0.10$) shows the positive slope of the susceptibility $\chi(T)$, which is a signature of the pseudogap effect [17]. Moreover, the estimated fictitious maximal critical temperature $T_{cmax}^0 \sim 24$ K is still lower than expected for single-layer HTC compounds and does not allow to explain the differences with the maximal critical temperature in the other single-layer Tl-2201 ($T_{cmax}\sim85$ K) [21] and Hg-1201 ($T_{cmax}\sim97$ K) [22] compounds.

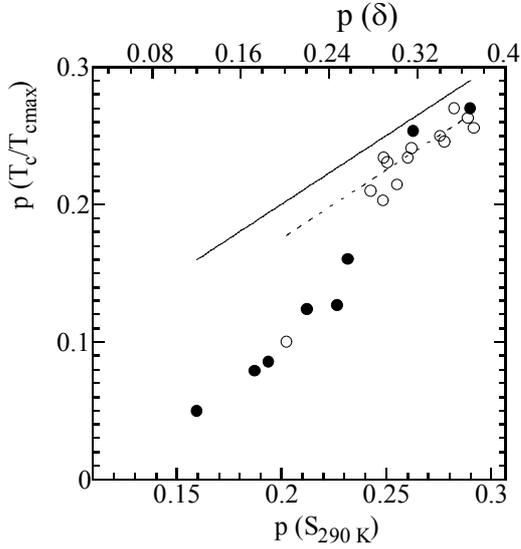
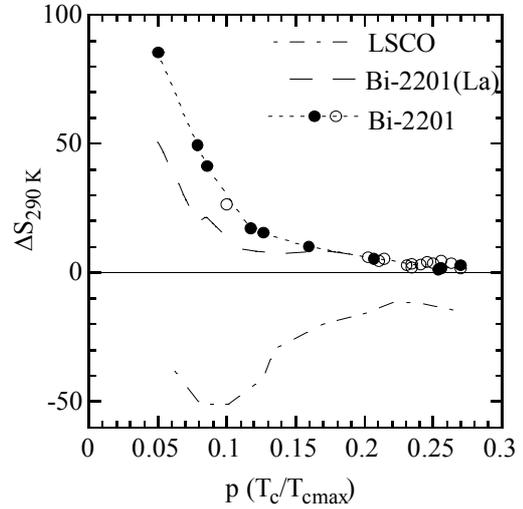

Fig. 4                                    Fig. 5

Fig. 4. – The hole number $p(T_c/T_{cmax})$ as a function of $p(S_{290K})$. Solid line shows $p(T_c/T_{cmax})=p(S_{290K})$. The agreement within 10% (dotted line) is observed for oxygen excess $\delta$ between $0.14<\delta<0.18$. In the top axis, hole number estimated from average valence of Cu, $p=2\delta$, assuming 3+ for the bismuth and 2+ for the strontium.

Fig. 5. – Discrepancies between empirical OCT behaviour and experimental data: $\Delta S_{290K}= S_{290K}-S_{290K}^{OCT}$ for given $p(T_c/T_{cmax})$ in the case of Bi-2201 (circles), Bi(La)-2201 (dashed line) [3] and LSCO (dashed-doted line) [2] family. Solid line represents OCT behaviour.

With the goal of analysing the second possibility, a somehow reduced measured thermoelectric power from the real one without affecting $T_c$, the difference between the measured $S_{290K}$ and empirical behaviour $S_{290K}^{OCT}$ is plotted as a function of $p(T_c/T_{cmax})$ (circles). The same analysis is done in the case of Bi(La)-2201 data from Ref.[3] (dashed line) and LSCO data from Ref.[2] (dashed-dotted line). As it was seen before the convergence of $S_{290K}$ for single-layer Bi-2201 family to OCT behaviour (solid line) is clearly indicated. Approaching weak hole numbers, $\Delta S_{290K}$ increases in an approximately exponential way. The LSCO compound shows completely opposite behaviour with negative $\Delta S_{290K}$. The convergence of $S_{290K}$ to $S_{290K}^{OCT}$ is signalled in the strongly underdoped nonsuperconducting behaviour [2], whereas a saturation of $S_{290K}$ to around ~10 μV/K is present in the overdoped region. Within the frame of the drag model [18,4], the positive thermopower values of LSCO could be explained with the anomalously large contribution of drag, arising from the existing excitation of carriers. In the case of Bi-2201 family, the drag contribution is diminished down to zero leaving only negative diffusion contribution in the strongly overdoped region ($\delta$>0.17) [15]. On the other hand, the anomalously negative thermopower in Bi-2201 is possibly related to the underlying electronic structure, giving rise to the non-classical Fermi-liquid diffusion [15,23].

To avoid the above discrepancy, Ando *et al.* [3] used the universality of Hall coefficient measurements in cuprates in order to estimate the hole number $p$, relaying on the well-determined $p$ in LSCO family. The predicted fictitious $T_c^0$ values are bigger than the measured one $T_c$, even in the overdoped region where our present results indicate a convergence to the universal behaviour. On the other hand, the hole number taken from the average valence of Cu (top axis in fig.4), is also overestimated, in the comparison with other hole-number determinations. Finally, the disagreement between $p(T_c/T_{cmax})$ [7] and $p(S_{290K})$ [8] can be avoided using a modified phenomenological relation $T_c/T_{cmax}=1-Z(p-p_{opt})^2$ with $Z$ and $p_{opt}$ different than the usual values of 82.6 and 0.16 respectively [24]. However, the difference with other single-layer Tl-2201 and Hg-1201 cuprates remains unclear.

All the above discussed approaches reveal the complexity of the Bi-2201 family. The increase of the residual resistivity $\rho_0$ and the appearance of a localisation effect in Bi-2201 [11] indicate the presence of disorder which could diminish the critical temperature. However, a reduced critical temperature without influence on the thermopower can not explain the observed untypical behaviour from that of other single-layer compounds. At the same time, the underlying electronic structure, giving rise to the robust negative values of thermopower, might be related to the structural distortions found in $Bi_2Sr_2CuO_{6+\delta}$ [25,26]. Moreover, the substitution of Sr by La reduces these distortions [25] and approaches Bi(La)-2201 to the other cuprates (fig. 3). The observed distortions linked to charge transfer could also lead to the overestimated p values, as determined from the Cu valence. More work is necessary to clarify this point.

In summary, we show the detailed doping dependence of thermopower measurements on $Bi_2Sr_2CuO_{6+\delta}$ thin films and ceramic samples in the overall superconducting region. The results are in good agreement with the empirical OCT relation in the strongly overdoped region (0.14<$\delta$<0.18), while for $\delta$<0.14, a failure of the OCT behaviour is observed. This behaviour is similar to the doping dependence reported on Bi(La)-2201 samples in the overdoped region but it is very different in the underdoped part. The superconducting ($T_c$,$S_{290K}$) phase diagram appears to be extremely narrow, reflecting anomalous underdoped region in non-substituted Bi-2201, even more anomalous than the one found in La-doped Bi-2201.

We thank F. Rullier-Albenque and K. Behnia for stimulating discussion.